\documentclass[sigconf]{acmart}

\usepackage{tabularx,arydshln}
\usepackage{multirow}
\usepackage{soul}  
\usepackage{color} 
\usepackage{xspace}
\usepackage{listings}

\newcommand{\eg}{{\it e.g.,\ }}

\newcommand{\ie}{{\it i.e.,\ }}

\AtBeginDocument{%
  \providecommand\BibTeX{{%
    \normalfont B\kern-0.5em{\scshape i\kern-0.25em b}\kern-0.8em\TeX}}}

\copyrightyear{2024}
\acmYear{2024}
\setcopyright{rightsretained}
\acmConference[CHI '24]{Proceedings of the CHI Conference on Human Factors in Computing Systems}{May 11--16, 2024}{Honolulu, HI, USA}
\acmBooktitle{Proceedings of the CHI Conference on Human Factors in Computing Systems (CHI '24), May 11--16, 2024, Honolulu, HI, USA}
\acmDOI{10.1145/3613904.3642529}
\acmISBN{979-8-4007-0330-0/24/05}

\begin{document}

\title[Authors' Values and Attitudes Towards AI-bridged Scalable Personalization of Creative Language Arts]{Authors' Values and Attitudes Towards AI-bridged Scalable Personalization of Creative Language Arts}


\author{Taewook Kim}
\email{taewook@u.northwestern.edu}
\affiliation{%
 \institution{Northwestern University}
 \city{Evanston}
 \state{IL}
 \country{USA}
}

\author{Hyomin Han}
\email{hyomin@u.northwestern.edu}
\affiliation{%
 \institution{Northwestern University}
 \city{Evanston}
 \state{IL}
 \country{USA}
}

\author{Eytan Adar}
\email{eadar@umich.edu}
\affiliation{%
 \institution{University of Michigan}
 \city{Ann Arbor}
 \state{MI}
 \country{USA}
}

\author{Matthew Kay}
\email{mjskay@northwestern.edu}
\affiliation{%
 \institution{Northwestern University}
 \city{Evanston}
 \state{IL}
 \country{USA}
}

\author{John Joon Young Chung}
\email{jchung@midjourney.com}
\affiliation{%
 \institution{Midjourney}
 \city{San Francisco}
 \state{CA}
 \country{USA}
}






\begin{abstract}

Generative AI has the potential to create a new form of interactive media: AI-bridged creative language arts (CLA), which bridge the author and audience by personalizing the author's vision to the audience's context and taste at scale. However, it is unclear what the authors' values and attitudes would be regarding AI-bridged CLA. To identify these values and attitudes, we conducted an interview study with 18 authors across eight genres (\eg poetry, comics) by presenting speculative but realistic AI-bridged CLA scenarios. We identified three benefits derived from the dynamics between author, artifact, and audience: those that 1) authors get from the process, 2) audiences get from the artifact, and 3) authors get from the audience. We found how AI-bridged CLA would either promote or reduce these benefits, along with authors' concerns. We hope our investigation hints at how AI can provide intriguing experiences to CLA audiences while promoting authors' values.

\end{abstract}



\maketitle

\section{Introduction}

Advancing large language model (LLM) technologies can generate text based on the user's specifications. With their generative capabilities, LLMs have the potential to change \textit{creative language arts} (CLA)\footnote{We use the term ``creative language arts'' to indicate creative, aesthetic linguistic artifacts that can focus on aspects such as narrative craft, character development, literary tropes, and traditions of poetry. Creative language arts can take various media forms, such as writing, stage performance, and audio or video recording}. New LLM-based tools can provide authors with AI-generated suggestions based on the author's own writing or instructions~\cite{clark2018creative, lee2022coauthor}. LLMs can also change the audience's experiences. For example, LLMs can personalize the style of existing writing, providing a different language arts experience from the original~\cite{das2023balancing}.

With an LLM's ability to generate texts with user-given constraints, one possibility to expand CLA experiences is to use AI models as an intermediary between authors and audiences. That is, instead of simply supporting the author's writing process with LLMs~\cite{lee2022coauthor, gero2022sparks}, we explore the concept of AI-bridged CLA, 
where the author conveys their broad artistic vision to AI models and these models present personalized versions of the vision to the audience at scale, according to the audience's contexts and tastes. We ground the concept within the history of how technologies have transformed CLA. Personalization has been one strength of performing oral arts with a small audience, as the performer could adapt what they present to the audience's preferences and reactions~\cite{ong2002orality}. However, with technologies that can distribute CLA to larger audiences (\eg print media), these benefits may have been lost. That is, such technologies remove the role of intermediate performers and fix the content to one form. AI technologies, such as LLMs, might offer the benefits of both personalization and scalability, as they can take the author's vision and present it to the audience with personalizations they might enjoy. 

With AI-bridged CLA, we strive to understand the perception of human authors about this new type of CLA media. We ask: How should AI-bridged CLA be used to facilitate what human authors value while not hindering their creative practice? What do human authors expect from AI-bridged CLA?
We conducted a semi-structured online interview study with 18 authors from eight different CLA genres including poetry, novels, essays, screenplays, film scripts, pop-song lyrics, webcomics, and interactive fiction. Prior to the interview study, we designed 16 speculative scenarios (eight for lyric; eight for novel) of AI-bridged CLA as a slide deck (available in the supplementary material). We presented these scenarios to authors during the interviews to help them grasp the potential applications of AI-bridged CLA. We designed our study to reveal the potential benefits of AI-bridges CLA and the possible impacts on the dynamics between author and audience.


From the study, we found the dynamics between author, audience, and artifact and that the authors consider three types of benefits within these dynamics: 1) benefits authors get from the process (\eg joy, therapeutic effect, attachment), 2) benefits audiences get from the artifact (\eg empathy to the artifact, entertainment), and 3) benefits authors get from the audience (\eg resonation from the audience, monetary reward). The authors expected AI-bridged CLA to impact these dynamics and benefits. For example, authors with high attachment to their own artifacts worried that AI might distort their intentions in the artifact. These authors often value the audience's appreciation of the author's unique characteristics in the artifact and worry that AI might convey distorted content to audiences. 
On the other hand, many other authors were okay with AI transforming aspects that are not core to their attachment, and acknowledged that AI-bridged CLA can add benefits, such as making the artifact more understandable and entertaining to the audience while increasing empathy and monetary reward authors get from the audience. Along with these findings and the authors' concerns, we discuss potential approaches to support the creation (\autoref{dis:creation}) and distribution (\autoref{dis:safeguard}) of AI-bridged CLA and challenges \& opportunities of AI-bridged CLA  (\autoref{dis:profession}). We hope our study attracts more researchers to investigate how to leverage generative AI for scalable personalization in other types of content, such as music and videos.
\section{Background of Creative Language Arts: Towards AI-Bridged Ones}

\begin{figure}
    \centering
    \includegraphics[width=0.5\textwidth]{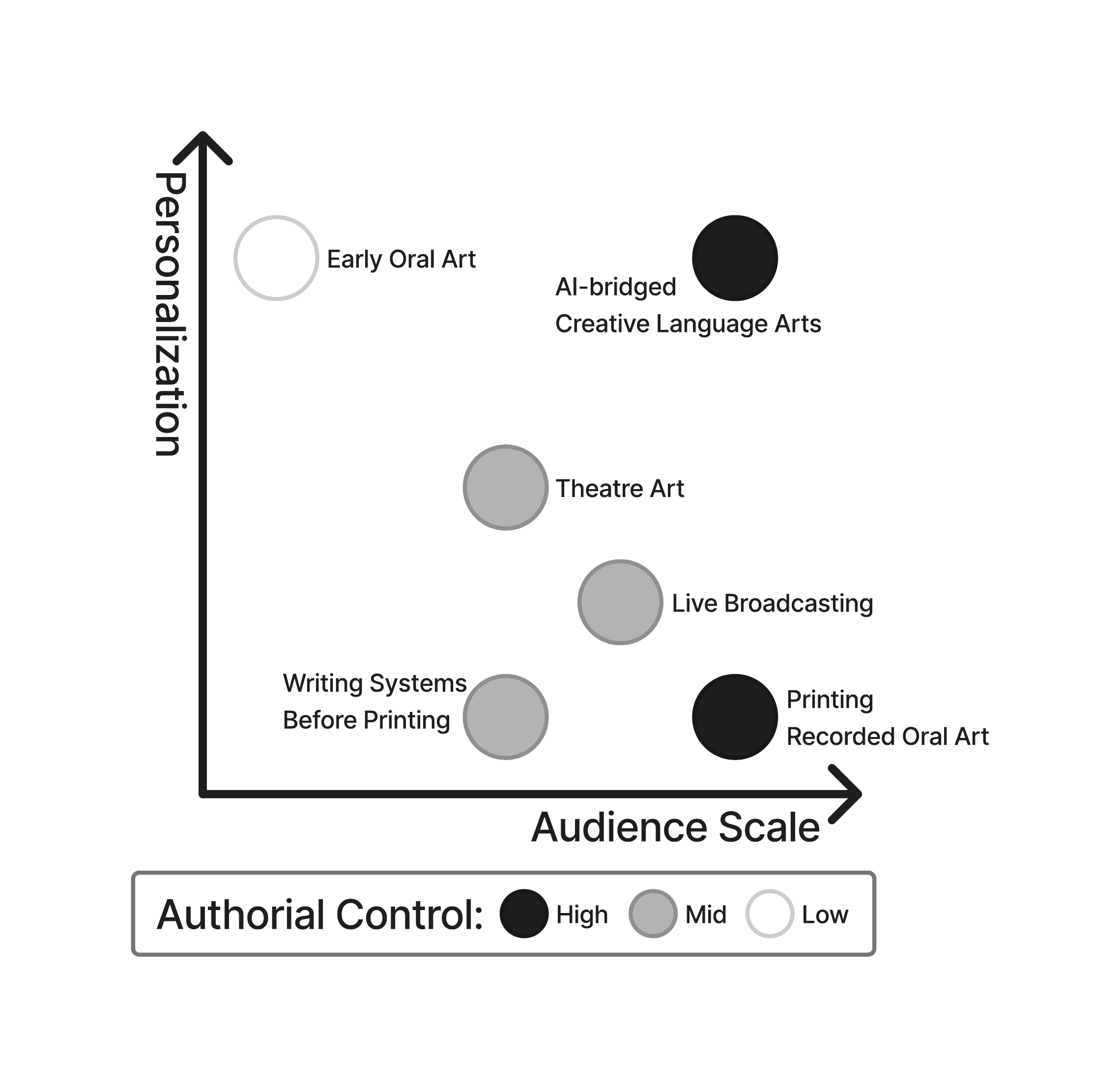}    
    \caption{The landscape of creative language arts media powered by different technologies, along dimensions of 1) personalization, 2) audience scale, and 3) authorial control. The dimension values of AI-bridged creative language arts are based on recent advances in AI but assume idealistic technology within the direction of advances.}
    \label{fig:CLA_landscape}
\end{figure}

To introduce the concept of AI-bridged CLA, we start with the history of how CLA changed with different technology-powered media. In this section, we first briefly discuss those media from pre-computer eras. Then, we expand to computer-powered CLA and introduce AI-bridged CLA. We derived the history and characteristics of different media from a collection of previous work. We present the overall landscape of these media in~\autoref{fig:CLA_landscape}. While there can be many different aspects regarding these media and technologies, we mainly focus on three aspects that would be most relevant to the author-audience relationship: 1) whether they are personalizable to the individual audience (personalization), 2) the scale of the reachable audience (audience scale), and 3) if the author maintains the control over the content (authorial control). Note that our concept of an author would be either an individual (\eg a sole novelist) or a collective team (\eg a movie production team) that makes decisions on creating a CLA piece.

\subsection{Creative Language Arts Before Computers}
\subsubsection{Early Oral Art}
Before writing systems were invented (\eg alphabets), performers conveyed narratives and poetry orally to others~\cite{ong2002orality}. As oral means themselves (\ie without any technologies) tend to be ephemeral, they would only reach limited audiences in the same time and space as the performer (\textbf{low audience scale}). Authorial control would be limited only with oral means, as the content can easily change from performer to performer (\textbf{low authorial control}). Moreover, the boundary between authors and performers would have been often unclear in the early oral arts. On the other hand, oral forms tend to be empathetic, participatory, and personalizable~\cite{ong2002orality}, as performers directly face the audience and can adapt the content to what the audience would be interested in (\textbf{high personalization}).

\subsubsection{Writing Systems Before Printing}

With recording tools like brushes, pens, papers, and inks, writing systems, such as the alphabet, introduced visual ways of conveying CLA instead of using oral means. They removed the need for human performers to convey linguistic content. People can distribute written CLA as publications to different places and times. However, with early writing systems, the size of the reachable audience would still be limited (\textbf{mid-audience scale}). For example, with manual handwriting, creating or copying writing still requires significant effort. Writing systems could also have enabled authorial control over CLA, as the authors can record the original texts in permanent forms with credits to themselves. However, as people used the mixture of oral and written media for a long time~\cite{ong2002orality}, the introduction of writing systems would not immediately have introduced the concept of authorial control (\textbf{mid-authorial control}). Another characteristic of writing is that they are much less personalizable than oral means (\textbf{low personalization}). As writing is recorded, there would be no performers directly interacting with audiences. Due to that, the authors should have considered the reader as a fixed persona and the content could not be contextualized or adapted to different readers as oral art did. 

\subsubsection{Printing}
Printing reinforced the characteristics of writing systems. As written content could be mechanically reproduced with little effort, printing greatly increased the audience scale of written CLA (\textbf{high audience scale}). With the massive distribution of printed writings, printing accelerated the transition from oral culture to writing culture. Printing also reinforced the notion of norm, credibility, correctness, and authorship, as authors can distribute a fixed content with minimal difference between copies~\cite{ong2002orality} (\textbf{high authorial control}). However, as printing copies the same content, it minimizes the chance of personalizing the content to individual audiences (\textbf{low personalization}). 

Note that variants of printed CLA allow a bit more extended personalization with high scalability. One type allows the reader to consume CLA non-linearly by listing multiple paths of consuming the content and allowing the audience to choose one of them. Choose-your-own-adventure style novels are examples of this type~\cite{montgomery2005journey, cook2016plotto}.  
With advanced distribution and publishing channels, printing different editions can be another variant. However, these still provide limited personalization compared to what oral arts could provide.

\subsubsection{Theatre Art}

Technology also changed CLA in oral forms. With improved and large stages, oral art forms, such as poetry recitals or plays, could be performed for larger and larger audiences. However, the audience still needs to be present at the performance, which is the limiting factor in the audience scale (\textbf{mid-audience scale}). Theater art became less personalizable than previous oral art. As the audience size grew, the performers could not react to each audience’s context or taste. However, reacting to mass audiences is still possible, as the performer is present with the audience~\cite{neuringer1987psychodynamics}. Moreover, while the final format of theater arts is in oral form, creating these often involves written scripts, which provide the high-level direction of the performance. 
With the fixed script and the credit of the scriptwriter on it, the level of personalization to the performance would likely be less than when written scripts did not exist
(\textbf{mid-personalization}). On the other hand, such written forms of oral CLA could have increased authorial control compared to early oral arts, as those scripts would limit transformations between performances (\textbf{mid-authorial control}). 
Note that authorial control of oral arts with massive audiences often tends to be distributed among multiple collaborators (e.g., directors, scriptwriters) and our concept of authorial control indicates the collective ones that are decided collaboratively during the planning (i.e., before the performance and personalization). 

\subsubsection{Live Broadcasting}
Additional technological advances, such as radio, provided the oral arts access to an even larger audience scale through live broadcasts. Through live radio or television shows, live broadcast performers did not have to be in the same space as the audience (\textbf{high audience scale}). However, with limited face-to-face interactions with their audience, personalization became even more challenging~\cite{hamilton2014streaming} (\textbf{mid-personalization}). For live broadcasts, the script would still give high-level direction, but the existence of performers who can make changes would limit the authorial control unless the performers themselves are the authors (\textbf{mid-authorial control}). Again, due to the collaborative nature of oral arts media that faces massive audiences, authorial control would likely be a collective one, and we consider it as collective decisions made before the performance and personalization. 

\subsubsection{Recorded Oral Art}
Audio and video recording provided yet another shift within the oral arts. These served the role similar to writing systems and printings---the performance could be recorded and replicated an infinite number of times in the same form. It freed oral performance from time and space constraints, making the medium reachable to a larger audience (\textbf{high audience scale}). However, similar to writing systems and printings, the possibility of personalization is greatly reduced as the recording forces a fixed form~\cite{neuringer1987psychodynamics} (\textbf{low personalization}). Moreover, as the author can make decisions on the final form of the artifact presented to the audience, the authors could have high authorial controls compared to live settings where the performer can bring in changes that misalign with the author's intention (\textbf{high authorial control}).

\subsubsection{Summary of Creative Language Arts Before Computers}
Oral traditions have seen significant evolution due to various technological advances that allowed the form to be distributed to larger audiences, beyond the constraints of time and space. Moreover, authorial control was strengthened, as technologies like printing allowed language arts to maintain their forms as the original author intended. However, as a trade-off, the personalization of early oral art gradually disappeared or weakened. As these media forms are recorded, copied, or broadcasted, performers who could previously recognize the audience's tastes and reactions and provide personalization either lost their roles (\eg reading does not require performers) or were restricted due to the scale of the audience (\eg the historical equivalent of streamers who cannot react to all individual audiences). Yet, personalization has the potential for interesting effects in language arts, such as adapting to the audience's reactions or fitting the content to the audience's taste. However, with the technologies mentioned in this section (before computers), personalizable CLA remains in forms that can reach only a limited set of audiences (e.g., oral arts performed for a small audience).

\subsection{Creative Language Arts With Computers: Towards AI-Bridged Ones}

We discuss how computers shifted CLA practices and how recent AI technologies like LLMs would make AI-bridged CLA more feasible. We first provide background on how computer technologies including AI shaped our CLA experiences in fixed written media. Then, we give a description of how computers, including AI, facilitated another type of CLA media, which can personalize their content to the audience. We extend this discussion to the concept of AI-bridged CLA: where AI models can accurately recognize the author’s artistic vision with flexible personalization on how such vision is presented to the audience. 

\subsubsection{Using Computers for Fixed Written Media}

For producing CLA in fixed, written formats, there have been many computer-based tools. They span from word processors such as Microsoft Word or Google Docs to more specialized tools, including grammar corrector~\cite{leacock2022automated}, machine-learning-based thesaurus~\cite{gero2019how}, crowd-powered writing assistants~\cite{bernstein2010soylent, nebeling2016wearwrite, huang2020heteroglossia, kim2014ensemble, kim2017mechanicalnovel}, or tools for specific types of writings, such as stories~\cite{plottr, dramatica}, poems~\cite{gero2019metaphoria}, lyrics~\cite{watanabe2017lyrisys}, help requests~\cite{hui2018introassist}, journalism~\cite{maiden2018making}, mental support~\cite{peng2020exploring}, and even affectionate messages~\cite{kim2019love}.

Researchers and practitioners also investigated how we can leverage generative AI technologies to provide support in producing CLA in fixed forms. While generation of CLA has been explored with various technical backbones, including template-based approaches~\cite{colton2012fullface, goncalo2019copoetryme}, symbolic planning~\cite{lebowitz1983creating, meehan1977talespin, penberthy1992ucpop, riedl2010narrative, ware2014computational}, case-based reasoning~\cite{gervas2005story, perez2001mexica, riedl2009vignette, swanson2012say, turner1992minstrel}, or character simulation~\cite{cavazza2001characters, louchart2007building}, LLM technologies powered by transformer architecture~\cite{vaswani2017attention} have brought in a large leap in flexibility and accuracy of text generation, as these models could be ``prompted’’ to serve arbitrary natural language tasks~\cite{brown2020language, openai2023gpt4, touvron2023llama}. With these LLM capabilities, many writing tools have been introduced, and one type of tool is those that suggest text phrases to the user’s writing, which is often called human-AI co-writing~\cite{mirowski2023cowriting, Kreminski2022looseends, gero2022sparks, arnold2020predictive, yuan2022wordcraft}. Researchers studied how these LLM suggestions can change people’s writing and found that generated texts could spark new ideas~\cite{clark2018creative, gero2022sparks, calderwood2020novelists}, lower grammatical errors~\cite{lee2022coauthor}, and increase vocabulary diversity~\cite{lee2022coauthor}, while introducing cognitive challenge of integrating generated texts into the user’s writing~\cite{calderwood2020novelists, nikhil2022where}. Researchers also studied how specific designs of LLM, such as the number of suggestions~\cite{buschek2021impact} or allowing users to input instructional prompts or not~\cite{dang2023choice} can impact writing. Moreover, the researchers investigated the sense of ownership of writings with AI suggestions~\cite{draxler2023ai}. However, AI suggestions were not the only approach to support human writing. For example, generating a summarization of the user’s writing to provide an external viewpoint can be one approach~\cite{dang2022beyond}. Researchers also studied socio-technical aspects of how writers interact with LLM-powered writing tools~\cite{gero2023social} and how the writer's own value would collide with the use of AI~\cite{biermann2022from}.
Moreover, researchers introduced approaches to evaluate LLMs from the perspectives of interacting with human writers~\cite{lee2023evaluating}. While this large body of literature helps us understand how LLMs can support writing fixed artifacts, they did not study the use of LLMs for personalizable media. 

\subsubsection{Using Computers for Personalizable Creative Language Arts}
\label{sec:concept}

With computer technologies, the personalization of linguistic content has become more feasible. Personalization can take two forms~\cite{adar2017persa}: 1) recommending different content to users~\cite{chesnais1995fishwrap, gabrilovich2004newsjunkie, isinkaye2015recommendation}, or 2) transforming the content to the user~\cite{knutov2009ah}.
We scope our discussion of AI-bridged CLA to transforming the content to users, as LLMs would extend transformative characteristics of such media.

One early type of computer-powered personalized CLA is adaptive hypertext~\cite{knutov2009ah}, which links the user to different content based on their contexts or direct input. Interactive narrative~\cite{murray1997hamlet} is one specific type, that provides the reader with a set of pre-authored options during the narrative progression so that the audience can explore a set of multiple story plots. These are analogous to choose-your-own-adventure novels and have been popularized as forms of digital games, visual novels, or interactive drama. Interactive poetry, which engages user direct manipulation into the presentation of poetry, is another~\cite{taper}. 

By linking the user to different content based on contexts and inputs, adaptive hypertext introduces the potential of crafting personalizable CLA. At the same time, as hypertext authors create different paths of content by themselves, authorial control would be largely maintained. Moreover, as digitized adaptive hypertexts can be distributed through the internet, these contents can reach a large audience when the audience wants to. However, personalization of adaptive hypertext would still have limited flexibility. With adaptive hypertext, personalizing contents need to be pre-authored, and hence, the number of possible personalizations would be constrained to a finite set.

AI can open new opportunities for AI-bridged personalizable CLA, which conveys the author’s intentions to the audience at scale, with adequate transformations that align with the audience’s contexts and reactions~\cite{riedl2010scalable}. AI-bridged personalizable CLA extends the flexibility in personalizing the content, beyond the user selecting one of the options/links in hypertext. 
One thread of research explored approaches to recognize the user context, instead of getting their direct input. Often in digital games, researchers and practitioners explored approaches to dynamically change content elements based on various aspects~\cite{panagiotis2023game} such as user preferences~\cite{thue2007interactive, yu2012asequential, yu2013data, yu2014personalized}, and affective states~\cite{karpouzis2013user}. 
However, these approaches either required a lot of technical knowledge in authoring the content or had limited flexibility in personalization. For example, with planning-based approaches, AI could select which story path to follow, but each story plot element should have been manually authored. 

With extended intelligence, LLMs would allow more flexible experiences in authoring and personalization. These models can receive flexible input from audiences and authors, such as prompts~\cite{brown2020language} or preferences~\cite{ouyang2022training}. Moreover, with generative capabilities, LLMs do not restrict personalized outputs to a pre-authored set of contents. 
However, one concern with this extended flexibility is losing authorial control with the generation, similar to how those controls were absent with frequent transformations in early oral arts. However, with some approaches to align LLM generations to the authors’ intentions, such as prompting~\cite{brown2020language} or tuning with human feedback~\cite{ouyang2022training}, the authors can increase the probability of aligning generated content with the authors’ intentions. Moreover, from the authors’ perspectives, generative LLMs can lower the required efforts in personalizing content, as they do not have to pre-author all possible content options. The scale of the audience would be high, as they can be distributed digitally at the moment when the audience wants. That is, ideally, with advanced AI technologies, AI-bridged CLA could achieve all of the high personalization, audience scale, and authorial control (\autoref{fig:CLA_landscape}).

Researchers and practitioners explored leveraging extended AI capabilities in AI-bridged personalizable CLA. For example, AI dungeon was designed to allow users to enjoy AI-powered interactive fantasy stories with free-form text inputs~\cite{aidungeon2_2019}. Some allowed the design of artificial characters and world elements that can autonomously interact with the user~\cite{kreminski2020why, park2023generative}. While previous work shows the feasibility of personalizable CLA extending the audience experiences, less is known about what the author's expectations would be regarding these emerging types of media. Thus, we investigate what authors expect, desire, and imagine with AI-bridged CLA using speculative scenarios.

\begin{figure*}
    \centering
    \includegraphics[width=\textwidth]{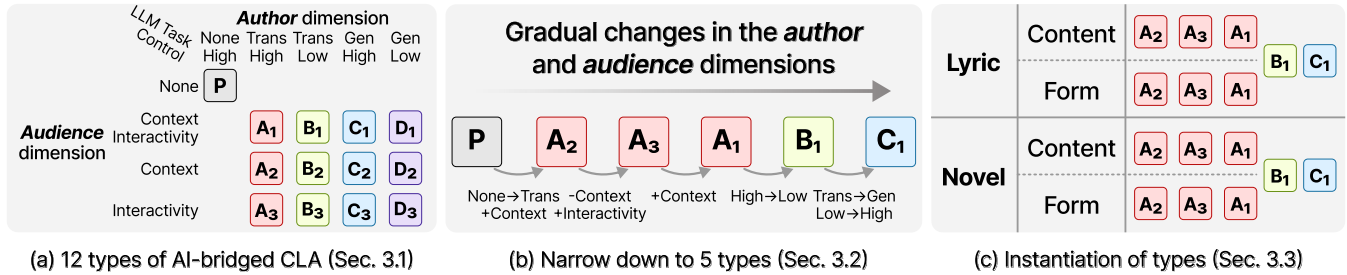}
    \caption{Overview of our scenario design process. \texttt{Trans} and \texttt{Gen} indicate transfer and generation, respectively. We first identified 12 types of AI-bridged CLA along author dimensions (LLM task, control) and audience dimensions (context, interactivity). We narrowed these into five types, gradually adding changes to the author's current practice ($P$). Then, we instantiated five types in genres of lyrics and novels while focusing AI transformations on content, form, or both.}
    \Description{There are three boxes. The first box is titled "(a) 12 types of AI-bridged CLA (Section 3.1). In the box, there are two axes, each named as author dimension and audience dimension. Under Author dimension, there are two subdimensions, "LLM Task" and "Control." There are five items under author dimension, "None High," "Trans High," "Trans Low," "Gen High," and "Gen Low." Under audience dimension, there are four items "None," "Context," "Interactivity," and "Context Interactivity." The space between these two dimensions are filled with individual items. The intersection of "None High" (author dimension) and "None" (audience dimension) is filled with "P". The space between other four author dimension items and other three audience dimension items are also filled with items. With author dimension items, individual intersecting items have prepending character "A," "B," "C," "D," respectively for "Trans High," "Trans Low," "Gen High," and "Gen Low." With audience dimension items, individual intersecting items have appending character "1," "2," "3," respectively for "Context," "Interactivity," and "Context Interactivity." The second box is titled "(b) Narrow down to 5 types (Section 3.2)." It has the upper caption "Gradual changes in the author and audience dimensions." Under the caption, there is a arrow heading right, and beneath it, there are individual intersecting items from (a), which are "P," "A2," "A3," "A1," "B1," and "C1." "P" is connecting to "A2" with arrow annotated "None-->Trans / +Context." "A2" is connecting to "A3" with arrow annotated "- Context / + Interactivity." "A3" is connecting to "A1" with arrow annotated "+Context." "A1" is connecting to "B1" with arrow annotated "High-->Low" "B1" is connecting to "C1" with arrow annotated "Trans-->Gen/Low-->High" The third box is titled "(c) Instantiation of types (Section 3.3)." It is divided into two dimensions in rows, "Lyric" and "Novel." Each are further separated to "Content" and "Form." In each row of "Content" and "Format" there are three items, "A2," "A3," and "A1." After that, the separation between "Content" and "Form" is gone, and there are two items "B1" and "C1."}
    \label{fig:scenario}
\end{figure*}

\section{Scenario Design}
\label{sec:scenario}

We interviewed CLA authors to investigate their expectations toward AI-bridged CLA (\autoref{sec:method}). However, they may have limited prior experience in AI-bridged CLA and related technologies. To help them imagine tangible examples of AI-bridged CLA, we developed several representative scenarios to use as a probe~\cite{hutchinson2003technology, renee2019hawkeye} when interviewing them. We characterized the speculative space with dimensions around two stakeholders: \textit{author} and \textit{audience}.

\subsection{Dimensions: Author and Audience}
\label{sub:dimensions}

\subsubsection{Author dimensions}
We first focused on two types of LLM-centered authoring tasks: \textit{transfer} and \textit{generation}. \textit{Transfer} tasks are those in which LLMs change parts of the author's original artifacts (\eg stylistic transfer~\cite{lee2021enhancing}). 
In contrast, \textit{generation} tasks are those in which the LLMs create parts of artifacts that did not exist yet (\eg story generation~\cite{chung2022talebrush} or poetry generation~\cite{ormazabal2022poelm, popescu-belis2022constrained}). In interacting with LLMs, authors can set different levels of authorial control: \textit{high} and \textit{low}. With \textit{high} control of a transfer task, for example, the author's story would ensure that the central events of the story, its background, and the main characters remain unchanged. With \textit{low} control, the transfer could change these elements. Generation with \textit{high} control would ensure that the produced content strictly followed the author's specifications. Based on the combination of \textit{LLM tasks} and \textit{levels of authorial control}, we can derive four combinations: Transfer/High ($A$), Transfer/Low ($B$), Generation/High ($C$), and Generation/Low ($D$). 

\subsubsection{Audience dimensions}
We speculated that AI-bridged CLA could vary in two main aspects for audiences: \textit{context} and \textit{interactivity}. Context refers to the audience's traits and situations, such as their background, preferences, personalities, device information, location, and time restrictions. Video streaming recommendations (\eg Netflix) are an example of using context to personalize content (without transforming the content itself). Interactivity reflects the amount of possible audience interaction with AI during the scalable personalization of CLA. AI Dungeon~\cite{aidungeon2_2019} is an example of incorporating interactivity, as the audience can intervene to input text while the AI unfolds the story. We excluded scenarios that used neither audience context nor interactivity, as personalization cannot happen if we do not have at least one (\ie there would be no input for personalization). From these, we can surface three combinations: Context/Interactivity ($S_1$: Y/Y, $S_2$: Y/N, and $S_3$: N/Y) ($S \in \{A, B, C, D\}$). In summary, we obtained 12 combinations of scenarios (4 author dimensions $\times$ 3 audience dimensions) (see~\autoref{tab:scenarios}).

\begin{table}
\caption{12 possible scenario types for AI-bridged scalable personalization of creative language arts.}
\label{tab:scenarios}
\centering

\begin{tabular}{l l c c c }
\toprule
    & \multicolumn{2}{c}{\textit{Author}} & \multicolumn{2}{c}{\textit{Audience}} \\ \cmidrule(lr){2-3} \cmidrule(lr){4-5}
    
    & LLM task & Control & Context & Interactivity\\
    \midrule
    $A_1$ & Transfer & High & Y & Y \\
    $A_2$ & Transfer & High & Y & N \\
    $A_3$ & Transfer & High & N & Y \\
    $B_1$ & Transfer & Low & Y & Y \\
    $B_2$ & Transfer & Low & Y & N \\
    $B_3$ & Transfer & Low & N & Y \\
    $C_1$ & Generation & High & Y & Y \\
    $C_2$ & Generation & High & Y & N \\
    $C_3$ & Generation & High & N & Y \\        
    $D_1$ & Generation & Low & Y & Y \\
    $D_2$ & Generation & Low & Y & N \\     
    $D_3$ & Generation & Low & N & Y \\
\bottomrule
\end{tabular}
\end{table}

\subsection{Narrowing Down Combinations}

\begin{figure*}
    \centering
    \includegraphics[width=0.75\textwidth]{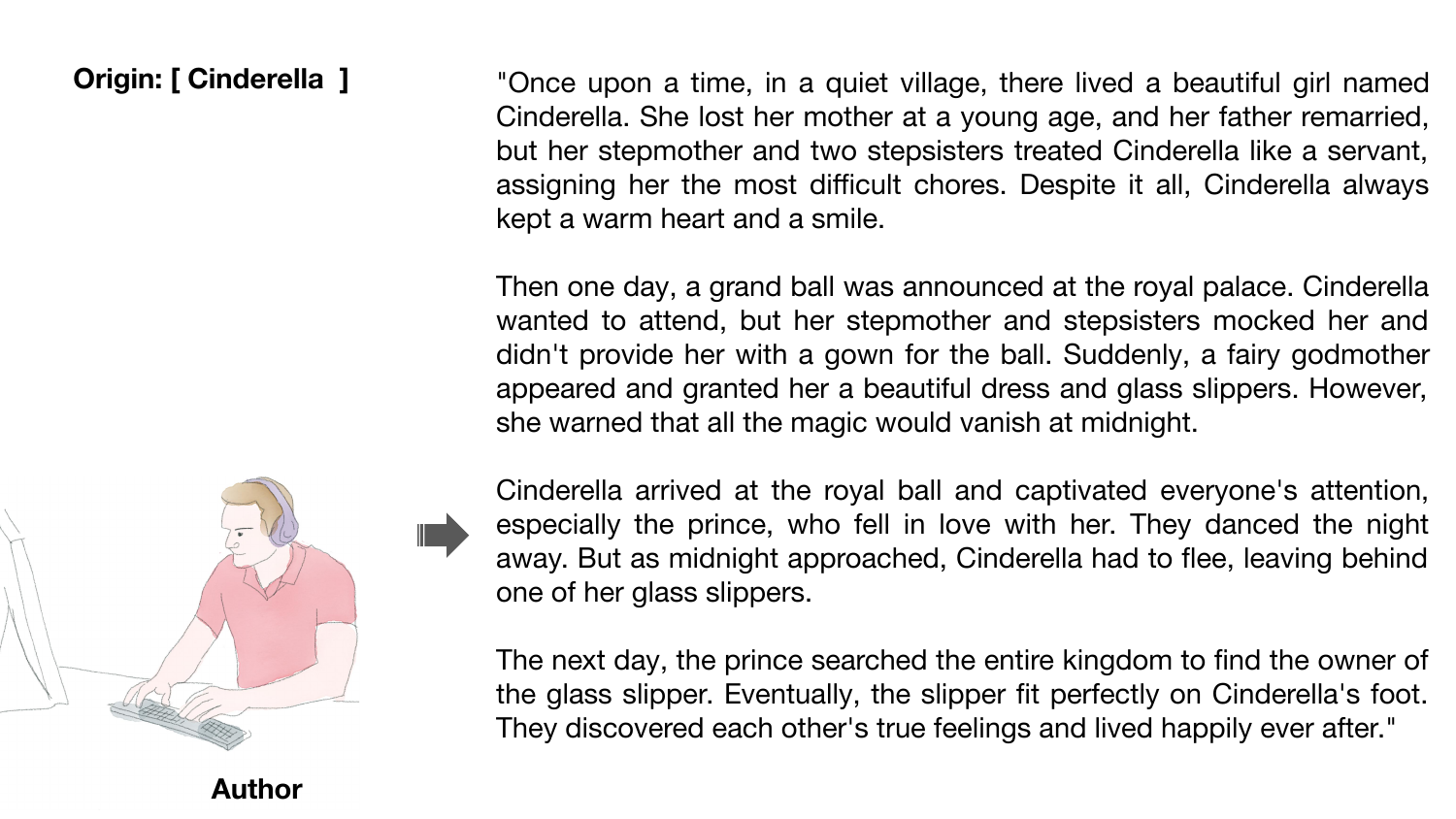}
    \caption{We used \textit{Cinderella} story to create AI-bridged CLA examples in novels. This is a translation from Korean to English.}
    \label{fig:original}
\end{figure*}

As 12 scenarios were infeasible to cover in a single interview, we restricted ourselves to five for the sessions. To select these, we first considered how authors' current writing practice ($P$) maps to the author and audience dimensions described in~\autoref{sub:dimensions}. In the author dimension, $P$ is \textit{without LLM} tasks but with \textit{high authorial control}, as authors have complete control over the output. In the audience dimension, $P$ does \textit{not} reflect the \textit{audience context}, such as audiences' background, and has \textit{no interactivity}. Thus, authors' current writing practice ($P$) is [n/a, high, N, N] (\autoref{fig:scenario} (a)).

Having $P$ as a baseline, we chose scenarios in a sequence with gradual changes in the author and audience dimensions. We picked $A_2$: [transfer, high, \textbf{Y}, N] as the first scenario because it is distinct from $P$ in one aspect (\ie context), except for the presence of an LLM task. Then we chose $A_3$: [transfer, high, N, \textbf{Y}] to examine distinctions in the audience dimensions. Next, we selected $A_1$: [transfer, high, \textbf{Y}, \textbf{Y}], in which both audience dimensions are reflected. We selected $B_1$: [transfer, \textbf{low}, Y, Y] as it shares all other aspects with $A_1$ except the authorial control. Lastly, we picked $C_1$: [\textbf{generation}, \textbf{high}, Y, Y], to show generation scenarios. In summary, we picked five types of scenarios ($A_2$, $A_3$, $A_1$, $B_1$, and $C_1$---see~\autoref{fig:scenario} (b)).

\subsection{Instantiation of Types}

\subsubsection{Two genres: lyrics and novels} For these five types, we brainstormed feasible and realistic scenarios for lyrics and novels, respectively, to show each version to relevant authors. As we wanted to ensure that all authors would know the original artifact, we selected \textit{Twinkle Twinkle Little Star} (lyrics) and \textit{Cinderella's story} (novel) as a stand-in for an author's original content (see~\autoref{fig:original}).

\subsubsection{Derivatives: content and form} We drew derivative scenarios based on two aspects --- \textit{content} and \textit{form} --- across all five combinations to ensure a broad coverage. In the literary critique, \textit{content} and \textit{form} are widely considered the core aspects of artifacts~\cite{amy1993generalizing, duncan1967dichotomy}. Content (\textit{what} it tells) includes properties such as theme, character, setting, plot, thought, etc., while form (\textit{how} it tells) considers syntax, narrative, versification, diction, imagery, etc.~\cite{duncan1967dichotomy}. We developed scenarios using these properties. However, it is not always clear whether certain properties affect solely content or form. For example, the change of narrative perspectives would affect the content. Hence, we differentiated the example scenarios based on the extent to which either \textit{content} or \textit{form} was relatively more modified.

\begin{figure*}
    \centering
    \includegraphics[width=0.85\textwidth]{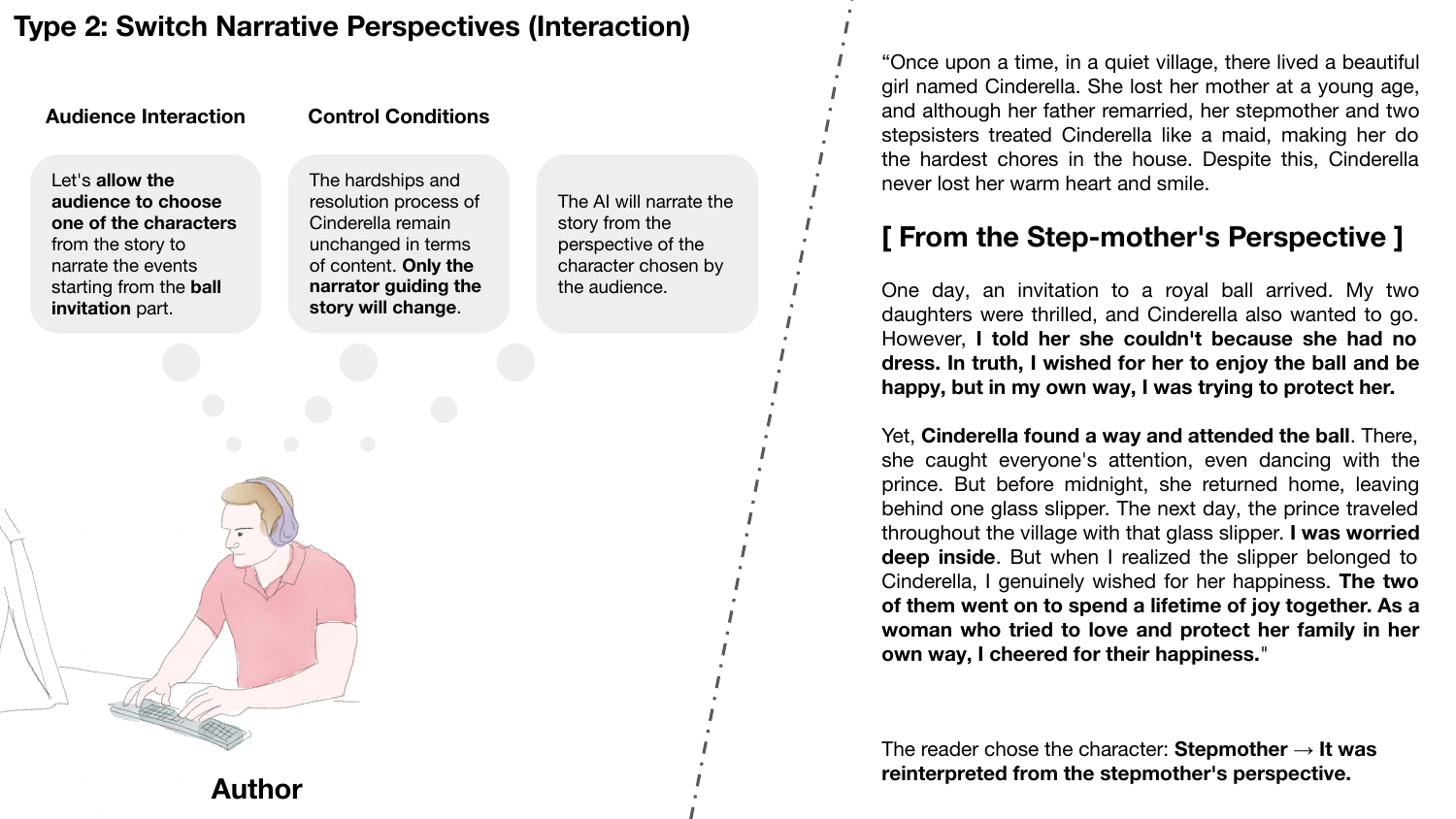}
    \caption{$A_3$-type ([transfer, high, N, Y]) derivative of \textit{Cinderella}. In this example, we explain to authors that audiences are allowed to interactively choose and switch narrative perspectives. Note that this is a translated version from Korean to English.}
    \label{fig:derivatives}
\end{figure*}

We finally instantiated five types for two genres (\ie lyric and novel) with two aspects (\ie content and form). However, we found that $B_1$: [transfer, low, Y, Y] (transfer with low control) and $C_1$: [generation, high, Y, Y] (generation without the original shape) make the distinction between content and form aspects more ambiguous. Therefore, for these two types ($B_1$ and $C_1$), we instantiated scenarios without differentiation of the content and form aspects. As a result, we projected 16 speculative scenarios of AI-bridged CLA (eight for lyric; eight for novel; see~\autoref{fig:scenario} (c)). We used GPT-4~\cite{openai2023gpt4} in designing these 16 scenarios. We manually edited some of the results from GPT-4 to improve the quality as a research probe. The entire scenario deck is available as supplementary material (an example in~\autoref{fig:derivatives}).

\section{Methodology}
\label{sec:method}
Our study aims to understand authors' views and expectations towards AI-bridged CLA. We interviewed CLA authors, presenting five speculative scenarios to guide our discussion (see supplementary material). Specifically, we looked into the authors' (a) writing practices (\textit{how} they write), (b) a ranking of the derived benefits of work (\textit{why} they write), (c) reactions to AI-bridged CLA, and how (a), (b), and (c) interact. To capture a broad perspective, we included authors from various CLA genres, ranging from traditional literature like poetry and novels to modern forms like (web)comics and interactive fiction, and considered authors at different career stages.

\subsection{Recruitment and Participants}

We recruited 18 authors (8 males and 10 females) working with eight different genres, including poetry, novels, essays, screenplays, film scripts, pop song lyrics, webcomics, and interactive fiction. To find these, we reached out to the CLA author community in South Korea, leveraging the first author's ties with them. We applied purposeful sampling to capture maximum variation~\cite{patton2002qualitative}. First, we selectively distributed our brochures to representative members across heterogeneous genres. We then asked them to recommend other authors who might be interested in our study, following a snowball sampling approach. While interviewing participants, we simultaneously recruited more authors to constantly evaluate our theory with new samples (see~\autoref{sec:analysis} for details).

\begin{table}

\caption{Background of participants. $\dag$ Int.Fiction denotes interactive fiction. *We specified the experience with LLM; the former refers to any experience with LLM, while the latter denotes the use of LLM in their creative writing process.}
\label{tab:participants}

\centering
\begin{tabular}{l l c r r c}
\toprule

    {\small\textit{\#}}
    & {\small \textit{Genre}}
    & {\small \textit{Age}}
    & {\small \textit{Training}}
    & {\small \textit{Experience}}
    & {\small \textit{Exp. w/ LLM*}}
    \\
    \midrule
    
    P1 & Poetry & 57 & Informal & 21 years & Y / N \\
    P2 & Poetry & 59 & Informal & 30 years & N / N \\
    P3 & Poetry & 57 & Informal & 31 years & Y / N \\
    P4 & Poetry & 48 &  Formal  & 15 years & Y / Y \\
    P5 & Novel & 67 & Informal & 41 years & N / N \\
    P6 & Novel & 45 & Formal & 21 years & N / N \\
    P7 & Novel & 31 & Formal & 10 years & Y / N \\
    P8 & Novel & 39 & Formal & 1 year & Y / N \\
    P9 & Essay & 61 & Informal & 15 years & N / N \\
    P10 & Essay & 28 & Formal & 3 years & Y / N \\
    P11 & Essay & 58 & Informal & 10 years & Y / N \\
    P12 & Screenplay & 54 & Formal & 23 years & Y / Y \\
    P13 & Film scripts & 43 & Formal & 12 years & Y / N \\
    P14 & Song lyrics & 44 & Informal & 15 years & Y / N \\
    P15 & Comics & 29 & Informal & 3 years & Y / Y \\
    P16 & Comics & 35 & Informal & 1 year & Y / Y \\
    P17 & Comics & 42 & Informal  & 17 years & Y / Y \\
    P18 & Int.Fiction$\dag$ & 33 & Informal & 3 years & Y / Y \\ 
\bottomrule
\end{tabular}
\end{table}

We tabulated details of our participants in~\autoref{tab:participants}. All participants had officially debuted as CLA writers and are currently active. Their experience ranged from 1 year to 41 years. We counted the participants' experience based on the year their first work was published. We filtered out those who have less than a year of experience in their active genre. All participants confirmed having a substantial history of publications within their respective genres. Some of them are active in more than one genre. For example, P5 made her first debut as a novelist, then expanded to poetry. We requested participants to specify their primary genre. During the interview, we instructed them to refer to this primary genre when answering questions.

We asked participants about their background. First, we asked them what training they received to become a CLA author (see the fourth column in~\autoref{tab:participants}). We indicate those who earned relevant higher degrees specifically in creative writing (\eg B.A., M.A., M.F.A., or Ph.D.) as formal training. We did not label participants (\eg P2, P3, P5) having higher degrees in literature as formal training because they claimed that their degree programs were academic (\eg analysis of CLA), with nothing to do with creative writing practices. The most common types of informal training are self-study and group study. Furthermore, we asked participants if they had previous experience with LLMs. We specified whether their experiences with LLM were part of the artifact creation process. Even if someone had never employed LLM for their literary creation process, those who used it for inspiration or information search were classified as having used LLM in their creative process.

\subsection{Interview Protocol}

Using slides of the five speculative scenarios, we conducted semi-structured individual interviews in Korean via Zoom with CLA authors from eight different genres. Although we did not require video, all but one (P10) had cameras on for the interview. With the consent of the participants, we began recording the interview for data analysis using theoretical coding (see~\autoref{sec:analysis}). The interviews took approximately an hour on average (M: 68.3 minutes, SD: 11.9 minutes). We paid each participant a virtual gift card in the amount of 70,000 KRW which is the equivalent amount of \$55 (USD) as compensation for their time.

At the beginning of the interview, we gave the participants a brief overview of the interview process. Before showing them our example scenarios designed in~\autoref{sec:scenario}, we first asked questions about (a) their writing practices (\textit{how} they write) and (b) their prioritized value of work (\textit{why} they write) for about 25 minutes. Then we introduced the five scenarios slides by screen sharing to see (c) their reactions to AI-bridged scenarios of scalable personalization. In some cases, participants were completely unfamiliar with LLMs. For example, P9 had never heard about LLMs and no understanding of their capabilities. Therefore, we explained each scenario one by one to ensure that they understood the concept of AI-bridged CLA. We also asked participants to imagine scenarios where their own artifacts are applied. We informed the participants that they could interrupt us at any time if they had any questions since the ideas introduced could be unfamiliar to them. For (c), we specifically asked questions like --- \textit{What do you like/dislike about this scenario? What do you wish to have (if possible) in this scenario? What kind of elements do you want to control to achieve your wish? In what contexts do you expect to have this scenario? What are the potential cases of misuse? Do you have other concerns?} This study protocol was reviewed and approved by the Institutional Review Board (IRB).

\subsection{Analysis}
\label{sec:analysis}

\begin{figure}
    \centering
    \includegraphics[width=0.44\textwidth]{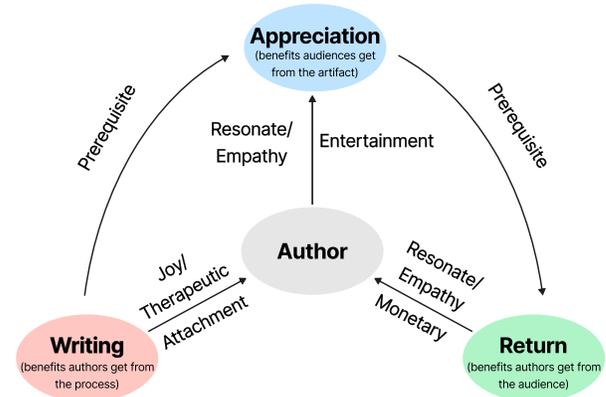}
    \caption{Benefit dynamics from the authors' perspective.}
    \label{fig:authors}
\end{figure}

We transcribed interviews using automatic transcribing software as soon as we finished each interview. Two authors who are bilingual in Korean and English translated them into English. We analyzed interview transcripts using theoretical coding that includes open coding, axial coding, and selective coding~\cite{muller2012kogan}. For the open coding process, two authors independently read the transcript, found quotes relevant to the research focus, and placed them on cards with labels (\ie low-level codes) in English. Then we collaboratively reviewed these low-level codes to resolve conflicts. We conducted this open coding once each interview was finished. We started the axial coding process when we finished about six interviews of three different genres (poetry, novels, and essays). In the axial coding process, we investigated relationships between low-level codes. Similar quotes were merged into high-level categories, and different ones were compared. These high-level categories resulted in: i) the benefits authors receive from the \textit{process}, ii) benefits audiences get from the \textit{artifact}, and iii) benefits authors get from the \textit{audience}. Through analysis, we identified the dynamics of these benefits from the authors' perspective (see~\autoref{fig:authors}). We repeated this analysis process concurrently while interviews were ongoing, to continue recruitment until diversity saturation was reached. We iteratively tested and refined our intermediate theories on the new data we collected. We regularly met to discuss, compare, and reshape codes and resolve disagreements through discussions. Finally, we derived two diagrams about (A) the dynamics of author-audience without AI, and (B) how AI-bridged CLA would influence these dynamics, as illustrated in~\autoref{fig:dynamics}.
\section{Findings: Authors' Values and Attitudes}

\autoref{fig:dynamics} summarizes the author-audience dynamics identified by our interview analysis. We found three types of derived values: i) benefits authors receive from the \textit{process}, ii) benefits audiences receive from the \textit{artifact}, and iii) benefits authors receive from the \textit{audience}. In addition, the authors showed a spectrum of reactions to AI-bridged CLA. These reactions are tied to values that the authors prioritize within the author-audience dynamics. We explain both the details of the author-audience dynamics without AI (\autoref{sub:dynamics_noAI}) and with (\autoref{sub:dynamics_AI}). Lastly, we introduce the author's concerns on the AI-bridged CLA in~\autoref{sub:concerns}.

\subsection{Author-Audience Dynamics without AI}
\label{sub:dynamics_noAI}

Our participants look for diverse values within their profession. What they prioritize varies as well. We categorized these benefits based on where they come from (see the colored legends in~\autoref{fig:authors}). Note that the authors tended to prioritize some---but often more than one---potential benefits. They all acknowledged and emphasized that they aim to achieve multiple benefits through CLA. We highlight those instances where authors notably prioritized certain derived benefits over others.

\subsubsection{Benefits The Author Gets from The Process}
For some authors, the benefit derived from the artifact creation process is highly important. These authors often find \textbf{therapeutic} (\eg P8, P9) and \textbf{joyful} (\eg P4, P9) experiences during the process of creating artifacts. Although they acknowledged other positive benefits, these authors highlighted the process-derived benefits. For example,

\begin{figure*}
    \centering
    \includegraphics[width=\textwidth]{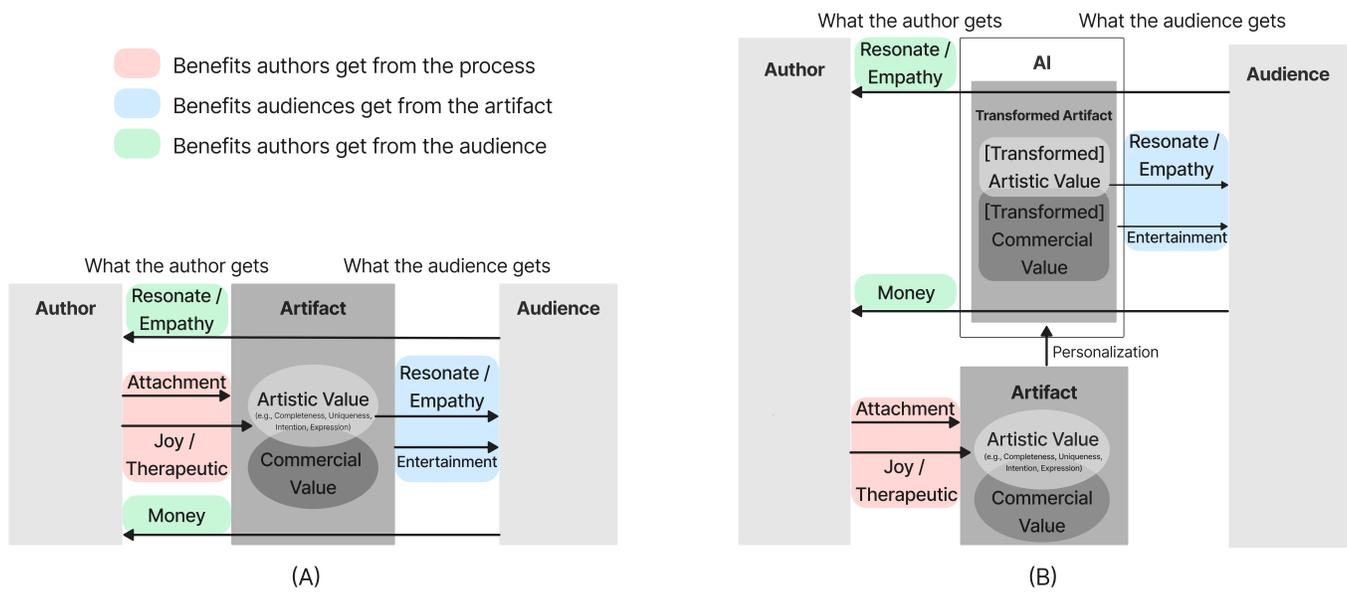}
    \caption{(A) Author-audience dynamics without AI. (B) Author-audience dynamics with AI-bridged CLA. Authors can find values from the process of creating the artifact (red). Authors infuse certain values into the artifacts in the hope that audiences would find these values from artifacts (blue). Lastly, authors might receive values from the audience (green).}
    \label{fig:dynamics}
\end{figure*}

\begin{quote}
P8: ``Writing has a therapeutic value for me, which seems to be the most important. Through writing, I reflect on my life and introspect. Such self-reflection helps me heal myself, understand myself better, and in turn, understand others as well.'' 
\end{quote}

Some authors emphasized the imbuing or realization of their own intrinsic artistic values in their artifacts, feeling \textbf{attachment} to those (\eg P1, P12, P17). They were particularly driven to increase these aspects (\eg uniqueness, completeness, infused intentions) within artifacts. For example,

\begin{quote}
P1: Literary value and uniqueness are the ones I most care about. Hence, if someone changes even a single letter, that work is no longer mine. I consider it a blemish on my originality.
\end{quote}

\subsubsection{Benefits The Audience Gets from The Artifact}
There are authors who emphasized the benefit that audiences get from the artifact as a motivating factor. The authors infuse values into the artifact, often hoping that those will reach the audience (\autoref{fig:dynamics} (A)). The authors were split on the nature of this goal. Some are keen on creating artifacts that audiences can deeply \textbf{empathize and resonate} with (\eg P2, P7, P10, P14), while others are more interested in creating artifacts that can provide pleasure and \textbf{entertainment} (\eg P6, P13, P15, P16, P18). Both are similar in that they desire their artifacts to hypothetically elicit certain reactions (\eg entertainment, empathy) from audiences. In other words, when certain properties are infused with the authors' intention, authors expect that their artifacts could evoke such reactions from audiences.


\begin{quote}
P10: ``I strive to ensure that my essays don't end as self-confessional stories, but rather as writings that can resonate with the reader's heart. I want to write pieces that allow readers to have their own realizations.'' 
\end{quote}
\begin{quote}
P16: ``I value entertainment and coherence. I eliminate content and scenes that seem uncomfortable to audiences. For example, in the case of the protagonist, even if they are in danger, I make sure it gets resolved in a short amount of time.'' 
\end{quote}

When making an impact on audiences through artifacts, authors wanted to correctly and effectively convey their intended vision within the artifacts to their audience. The authors described various strategies to achieve this. For example, some authors added universal objects (P4) or relatable elements (P7, P18) to provide familiarity. However, in the case that the author wants the audience to empathize, simply making every element familiar to the audience might not work as empathy presupposes a difference between the author and the audience. Hence, some authors mentioned that it was important to present familiar things in an unfamiliar way (\ie \textit{defamiliarization}~\cite{lawrence1984viktor}, but to the level that ensures the audience's understanding (P6, P10, P15). Others mentioned making catchphrases (P14) and receiving peer feedback (P13, P16) as ways to help audiences understand the authors' intentions effectively.

\subsubsection{Benefits The Author Gets from The Audience}

Some authors placed great importance on the benefits derived from the audience. Among these authors, some aim for occasional but tangible evidence (\eg social movement, a fan letter) of understanding and empathy from their audience (\eg P1, P2, P8, P10, P11), while others seek consistent monetary returns (\ie money) (\eg P15, P18). Authors could receive such returns from the audience, when their artifacts attract a broader audience, offer genuine entertainment, or resonate with the audience. That is to say, when values within the artifact are correctly and effectively delivered, authors can expect appreciation both in sentiment and financially, from their audience. For example,

\begin{quote}
P1: ``I occasionally receive emails from random fans, and I appreciate it. They said they've found themselves through my poetry. My work had been healing for them. I will keep it up to make such work.’’ 
\end{quote}
\begin{quote}
P18: ``I produce entirely commercial works. To be honest, creating work that can earn money is my top priority. My number one goal is to create content that can go viral and appeal to a broad audience.’’ 
\end{quote}

These authors make various efforts to achieve these benefits from the audience. First, the same effort to facilitate the audience to empathize/entertain with the author's work would also apply here, as the audience will more likely return empathy and monetary reward if they appreciate the artifact. In addition, some placed emphasis on creating their own unique artifact (\eg P1, P2) because they believe that such benefits from the audience can only be attained from work in the authors' unique tones. Others even use social media platforms such as personal blogs and Instagram to communicate with audiences (\eg P10, P11). They may also use these platforms to learn what is popular so as to create their own viral artifacts (\eg P18).

\subsection{Author-Audience Dynamics with AI}
\label{sub:dynamics_AI}
Our participants showed a spectrum of reactions to AI-bridged CLA. Their responses vary depending on benefits they prioritize within the author-audience dynamics. In~\autoref{fig:dynamics} (B), we illustrate how the author-audience dynamics change if AI is embedded. The most significant change is that the benefit that the audience gets from the \textit{artifact} and the benefit that the authors get from the \textit{audience} are mediated by the AI-transformed artifact---not the authors' completely original work. We emphasize this distinction in our analysis below. Note that the author-created ``artifact'' can both be the author's own piece to be transferred or the specification that is handed to AI for generation. Additionally, authors may individually have ambivalent or inconsistent attitudes toward AI-bridged CLA. 

\subsubsection{Prioritizing Author Benefits from The Process}
\label{sub:intention}
As AI-bridged CLA transforms the author's artifact, they presented varied responses depending on how much of an attachment they had to the created artifact. With high attachment, the authors increased their consideration of whether the transformed outputs of AI-bridged CLA preserved the authors' original intentions. However, there were differences among authors in terms of the intended aspects that were important. For example, some authors counted content and contextual properties (\eg backgrounds and personalities of characters in novels, materials in poetry) as a part of their intent (\eg P15). Others (\eg P1, P3, P17) considered not only the content but also the style and format of the writing (\eg poetic lineation, narrative perspectives). Therefore, some did not like changes in content and contextual properties, but were unconcerned about changes in format. Those who preferred not to change both content and format considered that every factor of the artifact represents the authors' intention. For example,

\begin{quote}
P1: ``The fact that the content of my poem can be consumed in any altered format is displeasing. If it has been changed in any way, it is not my work.''
\end{quote}
\begin{quote}
P3: ``I don't like these transformations. In CLA, messages/topics/themes should be aligned with the format. There's always the right format for each content.''
\end{quote}

Interestingly, while these authors showed a strong attachment to their artifacts, some mentioned that any changes through AI-bridged CLA could be acceptable, as long as their copyright and ownership were clearly ensured and the public would be fully aware of it (\eg P1, P17). 

In other cases, there were authors with very little attachment to their artifacts, who were fine with any changes in both content and format (\eg P2, P4). They even mentioned that they do not want to claim ownership of their artifacts. For example,

\begin{quote}
P2: ``I simply transcribe poetry that I receive from a spiritual entity. I have no interest in gaining fame from it.''
\end{quote}
\begin{quote}
P4: ``It would be nice if audiences could read my poems more freely and extensively. I don't want to claim my copyright and ownership for my work because I already found my happiness while writing my poems. I am happy with the act of writing itself.''
\end{quote}

However, both P2 and P4 worried about intentional misuse of their work. For example, P2 expressed concerns that AI could be misused to enforce standardization of individuals, like fascism, rather than facilitate diversity and personalization for individuals.

\subsubsection{Prioritizing Audience Benefits from The Artifact}
\label{sub:ai_artifact}

Authors imagined several potential benefits and harms to the audience from artifacts of AI-bridged CLA. First, many authors noted that the audience could be more entertained through interactive features and personalized content of AI-bridged CLA (\eg P3, P8, P12, P17). 

\begin{quote}
P12: ``Audience co-creating the story with AI, it can be very interesting and fun [for audiences].''
\end{quote}
\begin{quote}
P17: ``If I were a reader, I'd probably engage actively and be quite ambitious about it. From that point on, it might feel like a game to me.''
\end{quote}

P17 also mentioned that if the audience's interaction with AI-bridged CLA could have a significant impact on the story, it would be more fun and entertaining. Otherwise, audiences might not derive value from the interactive features.

Second, authors expected that personalized stories of AI-bridged CLA might facilitate audiences' appreciation and empathy with the content of the artifact (\eg P2, P10). They found that audiences could enjoy more immersive experiences from the artifact of AI-bridged CLA. For example, 

\begin{quote}
P2: ``The original purpose of poetry is to personalize it as audiences read. How to personalize it should be determined by individuals, not by the poet. If AI can help audiences achieve this, it would be wonderful.''
\end{quote}
\begin{quote}
P10: ``Switching perspectives and adapting my work pieces from each individual's own perspectives might enrich the appreciation of the essay much more effectively. It can enhance their level of immersion.''
\end{quote}

P8 introduced another benefit of personalized stories. It can increase the accessibility of the audience's understanding. She gave an example of a story about people smoking on airplanes, which was still allowed over 15 years ago. However, a young reader, having never lived in such an era, did not think it was a reflection of past society. The reader thought it was the author's imagination. The author of the story had never even considered that some readers might not understand or relate to such a true story.
\begin{quote}
P8: ``Young readers could struggle with understanding old pieces due to unfamiliar elements. In such cases, if AI-bridged CLA could personalize such unfamiliar elements into more relevant contemporary elements for the readers, that would be great. I believe it's essential for the sustainability of literature.''
\end{quote}

Lastly, other authors expected that audiences could be exposed to more diverse perspectives from artifacts of AI-bridged CLA (\eg P2, P5, P8, P10). They also hypothesized that individual audiences might be able to share their personalized content with each other and have extensive discussions about the original artifact.

\begin{quote}
P8: ``Transformation that shows another character's perspective is interesting as it could facilitate people to discuss what is not shown directly in the piece.''
\end{quote}

The authors also mentioned the potential harm to the audience. Specifically, they were concerned that some audiences could be exposed to inappropriate content (\eg age inappropriate) by AI-bridged CLA (\eg P5, P16). To prevent such incidents, they suggested that AI-bridged CLA should be able to suppress and facilitate different aspects of the content based on the audience profile. We compiled and discussed their concerns about uncertain benefits for audiences in~\autoref{sub:uncertain}.

\subsubsection{Prioritizing Author Benefits from The Audience}
\label{sub:monetary}
In this case, the authors expected a higher likelihood of \textit{return} (\autoref{fig:authors}) from the audience. Some predicted that authors could have increased financial benefits from AI-bridged CLA. These tools could potentially improve the commercial value of artifacts (\eg P9, P15, P18) and officialize audiences' secondary creation (\eg fan fiction) on an AI-bridged CLA platform (\eg P16, P18).

\begin{quote}
P16: ``Currently, derivative works created by readers tend to be shared on unofficial channels. It would be wonderful if, instead, anyone could officially produce derivatives via AI-bridged CLA platforms, and a part of the profits could be shared to the original authors.''
\end{quote}
\begin{quote}
P18: ``It feels like AI can actively assist with creating fan fiction. Even now, when you look at those who write fanfiction, they take creative liberties. I'd rather see more opportunities arising for the author to take character royalties through in-app purchases.''
\end{quote}

Others highlighted that authors would have more opportunities to receive empathy and reactions from audiences, as a consequence of increased exposure (\eg P1, P8, P15) and enhanced impression to audiences (\eg P10) by AI-bridged CLA.

However, some participants were concerned that their artifact may not resonate in the way they would intend if AI-bridged CLA is applied. For example, both P1 and P12 said that the uniqueness augmented in their artifact will be diminished by AI-modification. Then their audiences, who appreciated the author's voice, would no longer enjoy their work. 

\begin{quote}
P1: My readers read my works because they want to see the uniqueness of the author. If they can no longer see my voice in the poems, they won't want such poems. They'd have no reason to modify my poems through interactions. They probably wouldn't want to read my poems anymore.
\end{quote}

P1 and P12 view artifacts as a medium to express their unique individuality. Their primary goal is not mass appeal; instead, they aim for a niche audience that values and appreciates the authors' distinct voices and styles. They look for feedback and reactions from these exclusive audiences. Therefore, for these authors, it would seem to be a disruptive machine that could even blemish their unique voice as expressed through their artifacts. In contrast, P10 has a different perspective. For her, artifacts are more than just her self-reflection. She is eager to influence her audiences profoundly, using her essays as tools to encourage changes in their lives. Therefore, for P10, AI-bridged CLA could be an assistant to help achieve her goal by disseminating her work personalized to individuals at scale. The main difference between them would be that the former authors have a strong attachment to the artistic values they imbue in the artifact, and hope those to be accurately conveyed to the audience, while the latter prioritizes seeing the impact of the artifact on audiences.

\subsection{Concerns in AI Quality, Ethics, Audiences, and Profession}
\label{sub:concerns}

Despite many authors seeing potential benefits in AI-bridged CLA, our participants also raised several concerns. We categorized these by entities within the author-audience dynamics; AI itself (\autoref{sub:quality}), CLA artifacts (\autoref{sub:empathy}), audiences (\autoref{sub:uncertain}), and authors (\autoref{sub:profession}), respectively.

\subsubsection{Unreliable Quality of AI Performance}
\label{sub:quality}

Several participants raised concerns regarding the inconsistent quality of AI in handling artifacts. They questioned AI's capability to accurately identify and enhance the distinct characteristics of artifacts. The authors emphasized that the artistic merit and unique attributes of their artifacts are crucial for their value. However, they worried that the unreliable performance of AI might risk altering or even potentially misrepresenting the essence of their work.

\begin{quote}
P12: ``I have doubts regarding the ability of AI to produce high-quality stories. I'm concerned about whether AI could preserve an author's unique style and generate detailed aspects within the story.''
\end{quote}

Some authors corroborated this concern with an example. P7 said that LLM performance in direct \textit{telling} is comparable to humans, while it shows extremely bad quality in \textit{showing}, where P7 defined \textit{showing} as indirectly implying contextual information through descriptions. For example, `\textit{he was shocked}' is an example of \textit{telling}, while `\textit{he dropped his mug}' is an example of \textit{showing}. They explained:

\begin{quote}
P7: ``Audiences who have been trained in `showing' through long periods of writing or reading can often detect the shortcomings in ChatGPT's writing, feeling its lack and thirst, and they can easily identify that the text was written by ChatGPT.''
\end{quote}

Without higher-quality LLMs, they doubt that AI-bridged CLA could be successful.

\subsubsection{Misuse of AI-bridged CLA}
\label{sub:empathy}

Some authors raised a concern that AI-bridged CLA might accelerate the mass production of mundane artifacts. They noted that many authors overly reproduce similar patterns popular to the general audience. The authors were concerned that such practice might become easier and faster with AI---resulting in the audience seeing many pieces that are quite similar to each other (\eg P3, P5, P15).

\begin{quote}
P3: ``I'm afraid that AI would `reproduce' already similar content that is somewhat not very high-quality. It might bring detrimental effects on the overall quality of the art. Ultimately, it would be desirable if such democratization of creative arts could happen with the assurance of the overall quality.''
\end{quote}

The authors also worried about the potential misuse of AI-bridged CLA for distributing a uniform ideology to society. P2 viewed this as similar to how fascism leveraged mass media to spread its messages.

\begin{quote}
P2: ``What I am afraid of is the standardization by AI. If one uses AI to break individuals' uniqueness, and standardize us, then it becomes like `Big Brother'.''
\end{quote}

Many authors often take great pride and responsibility in their work (\eg P1, P14). In that sense, the deliberate misuse of AI-bridged CLA potentially entailing the mass production of low-quality work and dissemination of unintended messages could seriously tarnish the author's pride and reputation.

\subsubsection{Uncertain Benefits for Audiences}
\label{sub:uncertain}

Some authors were skeptical that there would be tangible benefits of AI-bridged CLA for audiences. They wondered why audiences would want to engage in the interactive features (\eg P1, P7, P17). P1 worried that AI-bridged CLA would hinder audiences from empathizing with the authors' intent correctly. Personalization would also negate the benefit of defamiliarization, which enables audiences to expand their experiences and knowledge (P1, P10). P17 claimed that some audiences would prefer a passive and relaxing experience over actively engaging in interactive features when they enjoy CLA.

\begin{quote}
P17: ``Audiences aged 30-40 read my comics to chill out. However, if AI continuously requires audiences to make decisions, it might be too exhausting. But teenagers, with their boundless energy, might enjoy it. I think preferences might vary depending on age.''
\end{quote}

P7 corroborated this concern, mentioning the decrease of interactive features on a video streaming platform (Netflix\footnote{\url{https://help.netflix.com/en/node/62526}}) possibly due to low audience usage.

\subsubsection{AI's Invasion of the Profession}
\label{sub:profession}
Authors had various concerns about their job security due to AI. While they acknowledged the inevitable impact of AI on their career, common reactions were about damage to pride (\eg P1, P8) and loss of opportunities for human authors (\eg P5, P7, P10).

\begin{quote}
P10: ``If AI reinterprets this world from various angles, our job opportunities would decrease.''
\end{quote}

Another thread was about copyright and ownership of artifacts (\eg P12, P14). All participants wanted to know how their copyright would be handled if this type of new media appeared in the future. Some questioned if they have to take the copyright and full responsibility of AI-bridged CLA content as well. It is at least clear that authors want to distinguish the boundaries of their ownership to protect themselves from AI-related issues.

\begin{quote}
P12: ``Copyright issue also would be relevant to the issue of responsibility.''
\end{quote}
\begin{quote}
P14: ``Responsibility is important as the society would react to created content, and if the generated content is inappropriate to the society, it can be problematic.''
\end{quote}

Some authors stated that authors should quickly accept the inevitable reality of AI invasion, and then proactively consider what they can do with AI. They shared a few ways that AI might have a positive impact, such as leveraging AI in their creative process (P2, P7, P12, P17) and intelligent matching between authors and audiences based on shared interests (P4).

\section{Discussion}

Authors value the connection that they establish with audiences through their artifacts. Authors receive benefits from the creation process (red in~\autoref{fig:dynamics}) and imbue values into the artifact, expecting audiences to also benefit (blue in~\autoref{fig:dynamics}). In return, authors hope to receive back benefits from the audience (green in~\autoref{fig:dynamics}). As AI intervenes within this dynamic, authors seek to maximize these benefits while not distorting the values they imbue into their artifacts. We discuss approaches to maximize the benefits of AI-bridged CLA by supporting the creation and distribution, challenges and opportunities of AI-bridged CLA, expanding AI-bridged scalable personalization to other media, and limitations.

\subsection{Supporting Creation of AI-bridged CLA through Authorial Controls}
\label{dis:creation}

Authors often have imbued values and intentions (\eg defamiliarization (P1, P10)) into their artifacts that they want to preserve in variations created by AI (\autoref{sub:intention}). Hence, AI-bridged CLA creation systems should enable authors to have some controls during the authors' creation process. First, AI-bridged CLA creation systems should enable authors to have flexible control for \textbf{constraining boundaries of variations}. For example, an author can categorically oppose any personalization that would create inappropriate versions (\eg P5, P16 in~\autoref{sub:ai_artifact}), such as turning a fairytale for kids into a pornographic narrative. In such cases, the system must be able to enforce strict restrictions to prevent these unwanted transformations (\eg using methods to prevent toxic degeneration of an LLM~\cite{gehman2020real}). 

However, some authors might struggle to envision what kind of variations would be possible. To facilitate the authors' speculation, AI-bridged CLA creation systems could \textbf{generate estimated variations}. Furthermore, such systems could provide valuable information on how different groups of audiences could react to these variations, through which authors could preview \textit{the benefits from the audience} (\eg AI-powered reaction simulation~\cite{joon2022social}). By engaging in several iterations of this reflection with the AI-generated estimations and their potential impact, authors might effectively preserve or even enhance the connection within author-audience dynamics with AI-bridged CLA.

\subsection{Supporting Distribution of AI-bridged CLA through Socio-technical Approaches}

\label{dis:safeguard}

Authors want AI-bridged CLA to strengthen \textit{the benefits authors get from the audience} (\autoref{sub:monetary}). They want to monitor and control unforeseen variations and leverage the reactions of audiences even after distribution. To achieve these needs, socio-technological approaches should be carefully considered in designing distribution platforms of AI-bridged CLA.

For example, distribution platforms could \textbf{record a stream of variations}. This could provide authors with a transparent view of how their artifacts are personalized and monetized at scale. Sharing records of variations on a blockchain could allow real-time monitoring and ensure transparency throughout the distribution process~\cite{frizzo2020blockchain, azzi2019power}. Furthermore, given the records of variations, the platforms may need to \textbf{prevent unauthorized deviations} beyond the original boundaries set by authors (\eg deliberate misuse of AI-bridged CLA (\autoref{sub:empathy})) even after the distribution. Watermarks for LLMs~\cite{kirchenbauer2023water} could be a technical option for this. For example, a platform could embed watermarks to personalized variations upon distribution, to restrict any further modification after distribution. In addition, the distribution platforms could provide community-based \textbf{interaction between authors and audiences}. For example, commentary interaction on video streaming platforms could allow authors to monitor the reactions of audiences. It could enable authors to learn what kind of variation audiences enjoy and discuss within the community. Authors can then incorporate what they have learned from these reactions into their future creations~\cite{guo2023online}.

However, any technological approach (\ie blockchain, watermarks) would likely be only temporarily effective and become obsolete as adversarial technologies also evolve. Thus, it is crucial to foster a more foundational discussion on cultivating social environments, \eg through education and policy, to develop a healthy culture around such technology.

\subsection{Challenges and Opportunities of AI-bridged CLA}

\label{dis:profession}

Authors (\eg P5, P7) are concerned about AI taking over their profession (\autoref{sub:profession}), which has also been clearly shown with the strike by the Writers Guild of America~\cite{wgastrike} against using AI to replace the writers' positions. This recalls the challenges traditional painters faced with the emergence of photography. As photography influenced a transformative shift in artistic expression~\cite{braun1977painting} (\eg impressionism~\cite{sweet2021before}, hyper-realism~\cite{bredekamp2006hyperrealism}), AI could similarly push the boundary of creative language arts. That is, authors might try to shun what AI models easily generate and create a new trend in creative language arts. However, this speculation assumes that AI would generate some types of writing more easily than others---potentially those more frequently found in the training data. As AI becomes more sophisticated, it could become `creative' in the sense of generating content beyond its training data distribution. In such cases, the traditional role of professional authors could diminish.

We believe, however, that humans will play a crucial role in the art world even with such advanced AI technologies. This is because arts are to be consumed by humans. With perpetual human roles as consumers and highly advanced AI technologies, prosumers---those who both create and consume content~\cite{Lang2021prosumers}---might be prevalent. This shift could democratize content creation. However, this also presents challenges, such as the potential to reinforce filter bubbles~\cite{bozdag2013bias}, where prosumers may predominantly create content that aligns with their own preferences and biases. Facilitating interactions between prosumers (\eg sharing personalization~\cite{bhuiyan2022other}) could be a solution to overcome such challenges. Moving towards prosumer media, AI-bridged CLA would be an intermediate format, as it can allow audiences to actively express ideas while still requiring the role of human authors. Prosumer media would, ultimately, merge the roles of authors and audiences in the dynamics of AI-bridged CLA, leaving two entities: prosumers and AI.

\subsection{AI-bridged Scalable Personalization in Other Creative Arts}

As generative AI has shown potential in diverse media formats such as audio~\cite{copet2023simple}, image~\cite{rombach2021highresolution}, and video~\cite{esser2023structure}, we can further explore the opportunities for AI-bridged scalable personalization in other types of creative media, like music and video arts. What if AI provides scalable personalization for music? For example, AI could automatically transform the mood of music to provide personalized experiences for listeners depending on context~\cite{Lu2019play}. Also, AI could synthesize the voices of listeners' favorite singers in playing any music to provide more personalized experiences for them~\cite{Liu2022diff}. Such transformations could be made at scale. However, similar to our study, we would need to first understand creators' perspectives on these technologies: how would music composers consider AI-bridged music, what kind of authorial controls do they want to have, how would their music-creating process be changed, and how would AI-bridged music influence the music composers' perceptions of their ownership~\cite{louie2022expressive}? Creators might have domain-specific concerns and desires for authorial control, which are understudied at the moment. Therefore, as we integrate AI into creative arts for scalable personalization, it is crucial to correctly understand the subtle differences in each medium.

\subsection{Limitations}

Our study has several limitations. First, we only interviewed authors. While incorporating the insights from the audiences would be informative for a richer understanding as our participants noted (\autoref{sub:uncertain}), we primarily focused on exploring the authors' perception of AI-bridged CLA. Therefore, future studies would be necessary. Second, our interview findings are based on hypothetical scenarios. Our speculative scenarios are just a few representative ones, but cannot entirely cover the wide spectrum of real cases. We may not know what the real applications would look like or how the authors would consider them. Thus, it might be necessary to investigate authors' and audiences' reactions to tangible applications of AI-bridged CLA in future studies. Finally, authors' reactions to AI-bridged CLA may have discrepancies depending on cultural or temporal contexts. Some authors wanted AI-bridged CLA to correctly preserve authors' original intent (\eg P15), while others argued that audiences should freely consume the content in their own ways (\eg P2). We consider that the degree of such variations can depend on cultural or temporal contexts. For instance, if an author and their audiences are in a society that encourages audiences to have a free and independent interpretation of the artifact rather than figuring out the author's intention and thoughts, the spectrum of reactions might differ.
\section{Conclusion}
Rapidly advancing generative AI technologies introduce the potential for AI-bridged CLA. These new types of interactive media can bridge the author and the audience by personalizing the author's vision to the audience's context. In this work, we investigate the author's perspective on AI-bridged CLA, the benefits they consider or anticipate, and their attitudes more broadly. For this purpose, using hypothetical scenarios of AI-bridged CLA, we conducted an interview study with 18 authors from eight genres. We found dynamics between the author, the audience, and the artifact. We also reflect on the benefits authors and audiences are likely to receive. We further identify how AI tools could add, reduce, or moderate the benefits that authors find important. Along with the authors' concerns, we discuss design implications for future AI-bridged CLA.


\begin{acks}

Special thanks to Mijung Kim for her support throughout this work. We thank Duri Long and the anonymous reviewers for their valuable feedback. Also, our appreciation goes to the participants for their time and sincere responses. This work was supported by the Northwestern Advanced Cognitive Science Fellowship.

\end{acks}

\bibliographystyle{ACM-Reference-Format}
\bibliography{references}


\end{document}